
%
%
%
%
%
\input phyzzx
%
%
\def\={\!=\!}
\def\-{\!\!-\!\!}
\def\+{\!\!+\!\!}

\def\shalf{{\textstyle{1\over2}}}   
\def\<{\left\langle}       
\def\>{\right\rangle}      

\def\d{\delta}
\def\D{\Delta}
\def\ga{\gamma}
\def\e{\epsilon}
\def\la{\lambda}
\def\lh{\hat\lambda}

\def\dl{{\cal D}\lambda}
\def\r{\rho}
\def\rh{\hat\rho}
\def\dr{{\cal D}\rho}

\def\x{\xi}

\def\N{{\textstyle{1\over N}}}
\def\iN{{\textstyle{i\over N}}}
\def\h{{\cal H}}
\def\T{{\cal T}}
\def\medint{{\textstyle\int}}
\def\medsum{{\textstyle\sum}}
\def\lhs{\hbox{\it l.h.s.}}
\def\rhs{\hbox{\it r.h.s.}}
%
%
%
\Ref\revival{ E. Br\'ezin and V.A. Kazakov \journal Phys. Lett.&B236 (90) 144;
   \nextline M.R. Douglas and S.H. Shenker \journal Nucl. Phys.&B335 (90) 635;
   \nextline D.J. Gross and A.A. Migdal\journal Phys. Rev. Lett.&64 (90)
127,717;
   \nextline E. Br\'ezin, M.R. Douglas, V.A. Kazakov, S.H. Shenker
   \journal Phys. Lett.&B237 (90) 43; \nextline
   C. Crnkovi\v c, P. Ginsparg and G. Moore \journal Phys. Lett.&B237 (90) 196;
   \nextline M.R. Douglas \journal Phys. Lett.&B238 (90) 176.}
\Ref\scaling{ I.K. Kostov and A. Krzywicki \journal Phys. Lett.&B187 (87) 149;
   \nextline I.K. Kostov and M.L. Mehta \journal Phys. Lett.&B189 (87) 118.}
\Ref\shenker{ S.H. Shenker, proceedings of the Carg\`ese workshop (May 1990);
   \nextline F. David, {\sl ibid}.}
\Ref\brezin{ E. Br\'ezin, C. Itzykson, G. Parisi and J.B. Zuber
   \journal Comm. Math. Phys.&59 (78) 35.}
\Ref\jurkiewicz{ J. Jurkiewicz \journal Phys. Lett.&B245 (90) 178.}
\Ref\bhanot{ G. Bhanot, G. Mandal and O. Narayan \journal Phys. Lett.&B251 (90)
   388.}
\Ref\jain{ K. Demeterfi, N. Deo, S. Jain and C.-I. Tan \journal Phys. Rev.&D42
   (90) 4105.}
\Ref\ray{ O. Lechtenfeld, R. Ray and A. Ray \journal Int. J. Mod. Phys.&A6 (91)
   4491.}
\Ref\bessis{ D. Bessis, C. Itzykson and J.B. Zuber
   \journal Adv. Appl. Math.&1 (80) 109.}
\Ref\banks{ T. Banks, M.R. Douglas, N. Seiberg and S.H. Shenker
   \journal Phys. Lett.&B238 (90) 279.}
\Ref\molinari{ L. Molinari \journal J. Phys.&A21 (88) 1;
   \nextline G.M Cicuta, L. Molinari and E. Montaldi \journal J. Phys.&A23 (90)
   L421;
   \nextline L. Molinari and E. Montaldi, INFN Milano preprint (1990).}
\Ref\tunneling{ O. Lechtenfeld \journal Int. J. Mod. Phys.&A7 (92) 2335.}
\Ref\semicl{ O. Lechtenfeld \journal Int. J. Mod. Phys.&A7 (92) 7097.}
\Ref\karabali{ D. Karabali and S. Sakita \journal Int. J. Mod. Phys.&A6 (91)
   5079.}
\Ref\menahem{ S. Ben-Menahem \journal Nucl. Phys.&B391 (93) 176.}
\Ref\titchmarsh{ E.C. Titchmarsh, \sl Introduction to the theory of Fourier
   integrals\rm\ (Oxford, 2nd ed. 1948) Chap.~V.}
\Ref\sohngen{ H. S\"ohngen \journal Math. Z.&45 (39) 245.}
\Ref\tricomi{ F.G. Tricomi \journal Quart. J. Math. Oxford Ser. (2)&2 (51)
   199.}
\Ref\suzuki{ M. Sasaki and H. Suzuki \journal Phys. Rev.&D43 (91) 4015.}
\Ref\senechal{ D. Senechal \journal Int. J. Mod. Phys.&A7 (92) 1491.}
%
%
\hfuzz=20pt
\nopubblock
\titlepage
\line{ITP-UH-25/93 \hfill}
\line{hep-th/9312124}
{\bf\title{Collective Field Theory \break
for D=0 Matrix Models~\foot{
\rm Invited Talk at the XXVII International Symposium in
Wendisch-Rietz, Germany, Sept. 7--11, 1993.}}}
\author{Olaf Lechtenfeld~\foot{
email address: \caps lechtenf@itp.uni-hannover.de}}
\address{Institut f\"ur Theoretische Physik, Universit\"at Hannover
\break Appelstra\ss e 2, D--30167 Hannover, Germany}
\vfil
\abstract
I investigate non-perturbative aspects of zero-dimensional matrix models.
Subtleties in the large-$N$ limit of the semiclassical picture are pointed out.
The tunneling of eigenvalues is seen to correspond to a chaotic sequence
of recursion coefficients determining the orthogonal polynomials.
\endpage
\chapternumber=0
\pagenumber=1
\sequentialequations
\noindent
{\bf{Introduction.}}

\noindent
Over the last four years, we have learned how to model two-dimensional
euclidean quantum gravity (with topological fluctuations) by hermitean
matrix models~[\revival], in the so-called double-scaling limit~[\scaling].
The latter entails sending the size~$N$ of the matrix~$M$ to infinity while
the couplings~$g$ in the matrix potential~$V$ approach a critical value~$g_c$
in such a way that some combination of $N$ and $(g\-g_c)$ is kept constant.
In the pure gravity case, one has
$$
V(M)\ =\ \shalf M^2 + gM^4 \quad,\qquad g_c=-{\textstyle{1\over48}}\quad.
\eqn\quarticpot $$
Interestingly, the critical matrix potential retains a local minimum but is
unbounded from below, and analytic continuation is to be performed
into a region where the partition function is ill-defined.
The result is either singular or develops an imaginary part, associated with
the instability towards the tunneling escape of individual matrix
eigenvalues~[\shenker].

The tunneling phenomenon is most transparent in a semiclassical treatment of
large-$N$ matrix models.
As a first step, the saddle-point analysis was pioneered in ref.~[\brezin].
To work with a well-defined theory, I will bound the potential~\quarticpot\
from below by changing its `large-$M$' behavior through a modification
$\d V(M)=\e M^6$, with a small, positive regulator~$\e$.
This turns out to change substantially the large-$N$ phase structure of the
model~[\jurkiewicz--\ray]; in particular, the critical line
$g_c\=g_3(\e)$ becomes {\it metastable\/} and invisible at finite~$N$.
Nevertheless, tunneling may now be studied directly, without the need for
analytic continuation.

Instrumental to the success of the random matrix model has been the
technology of orthogonal polynomials~[\bessis,\banks].
The saddle-point results, however, have not yet been completely understood
in this framework.
Only in the case of {\it degenerate\/} potential wells, a precise relation
between orthogonal polynomial recursion coefficients~$R_k$ on one side
and the (multi-band) classical eigenvalue density on the other side has been
established~[\molinari,\jain,\ray].

The more general connection between the two approaches involves an
interpretation of eigenvalue tunneling in terms of orthogonal polynomials.
This will be addressed in the third part of my talk, where I shall present
a surprising resolution of a puzzle mentioned earlier~[\jain,\ray,\tunneling].
The second part, following this introduction, will outline the semiclassical
approach.
Along the way, I will relate $N{=}\infty$~saddle points to airfoils, and
interpret eigenvalue tunneling in terms of the thawing of a frozen Dyson gas.
A short summary shall close my talk, which reviews some research of mine
conducted over the past three years and published for the most part in
refs.~[\ray,\tunneling,\semicl].
\vglue.2in
\goodbreak

\noindent
{\bf{The Semiclassical Approach.}}

\noindent
I begin by formulating a collective field theory for the zero-dimensional
hermitean one-matrix model at {\it finite\/}~$N$.  My starting point
is the partition function
$$
Z_N \ \propto\ \int\!d^{N^2}\!\! M \;e^{-N\tr V(M)}
\eqn\ZMdef $$
for an $N{\times}N$ random hermitean matrix ensemble, in a
potential~$V$.  Upon diagonalization $M=diag(x_i)$ this reduces to
$$
Z_N = e^{-N^2 F_N}\ :=\ \biggl[\prod_{i=1}^N \int\! dx_i\biggr] \; \exp\biggl\{
-N\sum_i V(x_i) + \sum_{i<j}\ln (x_i-x_j)^2 \biggr\} \quad,
\eqn\Zxdef $$
the partition function of a two-dimensional Coulomb gas of charges restricted
to a line, in an external potential $V(x)$ at temperature
$\beta^{-1}{=}1/N^2$.
I like to change variables from the matrix eigenvalues~$x_i$ to their
density distribution
$$
\r(x)\ :=\ \N\sum_{i=1}^N\d(x-x_i)\quad.
\eqn\rhodef $$
More precisely, I insert
$$
\eqalign{
1\ &=\ \int\dr\;\prod_x \d\bigl(\r(x)-\N\medsum_i \d(x\-x_i)\bigr)\cr
&=\ \int\!\!\!\!\int\dr\dl\;\exp\Bigl\{i\int\!dx\,\la(x)
\bigl[\r(x)-\N\medsum_i\d(x\-x_i)\bigr]\Bigr\} \cr}
\eqn\insertion $$
into \Zxdef\ and express the action in terms of the density,
$$
S_N^0[\r]\ =\ N^2\int\!\!dx\,\r(x)V(x) -\shalf
N^2\int\!\!\!\!\int\!\!dxdy \;\r(x)f(x-y)\r(y)+\shalf N\,f(0)\quad.
\eqn\Srho $$
The self-interaction had to be regulated by replacing $\,\ln z^2 \to
f(z)\,$ in eq.~\Zxdef, choosing some suitable, \ie\ symmetric and
bounded, function~$f$.

Following ref.~[\karabali] I am able to perform the integration
over~$x_i$,
$$
\eqalign{
Z_N\ &=\ \int\!\!\!\!\int\dr\dl\;\;e^{-S_N^0[\r]+i\int\!\la\,\r}\;
\biggl[\prod_{i=1}^N\int\!dx_i\biggr]\; e^{-\iN\sum_i\la(x_i)} \crr
&=\ \int\!\!\!\!\int\dr\dl\;\;e^{-S_N^0[\r]+i\int\!\la\,\r}\;
\biggl[\int\!dx\;e^{-\iN\la(x)}\biggr]^N \quad,\cr}
\eqn\Zrholambda $$
and arrive at an {\it exact\/} effective action
$$
S_N[\r,\la]\ =\ S_N^0[\r]-i\int\!dx\,\la(x)\r(x)-N\ln\int\!dx\,e^{-\iN\la(x)}
\eqn\Srholambda $$
which is not only nonlocal in the two real fields $\r$ and~$\la$ but
also non-polynomial in the latter.  Interestingly, the constant mode
of~$\la$ can be integrated out exactly to yield the constraint
$\d(\int\!\r-1)$ that was apparent already from the
definition~\rhodef.  However, I shall keep those modes in the measure
for the time being.  In principle, another constraint arises from the
positivity of~$\r$.  Perturbation theory about a strictly
positive~$\rh$, however, is insensitive to this restriction, and I
will therefore ignore it in the following.

My goal is to initiate a systematic semiclassical analysis of this
peculiar one-dimensional field theory.  To leading order in~$\hbar$ we
must determine the saddle-point configurations $(\rh,\lh)$, where the
action~\Srholambda\ is stationary.  The first variations yield
$$
\eqalign{
0\ &=\ N^2 V(x) - N^2\int\!dy\,f(x-y)\,\rh(y) -i\lh(x) \crr
0\ &=\ \rh(x) -e^{-\iN\lh(x)}\Big/\medint\!dy\,e^{-\iN\lh(y)} \cr}
\eqn\SPrholambda $$
where $\rh$ comes out to be properly normalized, $\int\rh=1$.  The
second equation determines~$\lh$ up to a constant,
$$
\lh(x)\ =\ iN\ln\rh(x)+\la_0 \quad,
\eqn\SPlambda $$
a result which may be inserted into the first equation.\foot{ The
first equation subsequently fixes the constant $\la_0$ for a given
solution $\rh$.} Differentiating with respect to~$x$ to remove
constants and deregulating the Coulomb repulsion I get\foot{
See also ref.~[\menahem].}
$$
\shalf V'(x)\ +\ {1\over2N}{\rh'\over\rh}(x)\ =
\ -\!\!\!\!\!\!\int \! {dy\over x-y}\;\rh(y)\ \equiv\ \pi\,\h_x[\rh]
\eqn\SPrhoc $$
where $-\!\!\!\!\!\int$ denotes Cauchy's principal value of the
integral.  The \rhs\ is known as the Hilbert transform \h\ (of
$\rh$) which has been thoroughly investigated~[\titchmarsh].
Together with normalization and positivity, this equation describes
the classical eigenvalue density for any finite~$N$.  It is noteworthy
that~\SPrhoc\ is not homogeneous in~$\N$, so its solution cannot be,
either.
At $N\=\infty$ the equation~\SPrhoc\ has been widely studied and
solved [\sohngen,\tricomi,\brezin], and it was learned~[\jurkiewicz,\tunneling]
that a {\it unique\/} solution extends to~$N<\infty$.  Unfortunately, the
equation is not easily solved for finite~$N$.  Even in the large-$N$
limit some care is required, as shown by the following.  For instance,
should I drop the $\rh'/\rh$-term since it is down by~$\N$?  A little
inspection reveals that such a step is in general not consistent with
the asymptotic large-$|x|$ behavior of the equation, which demands a
${1\over x}$ fall-off for the \lhs\ due to the normalization of~$\rh$.
In fact, the formal `solution'
$$
\rh_\e(x)\ \sim\
\exp\Bigl\{-N\bigl[V(x)-\medint\ln(x\-y)^2\,\rh_\e(y)\bigr]\Bigr\}
\ \mathrel{\mathop\approx^{|x|\to\infty}}\ x^{2N}\,e^{-NV(x)}
\eqn\mock $$
of equation~\SPrhoc\
shows that the $\rh'\over\rh$ term dominates the \lhs\ of eq.~\SPrhoc\ for
$|x|\gg1$ (unless $V\sim\ln x^2$), so that $\rh_\e={\cal O}(e^{-N})$
asymptotically.  Note that I have attached a subscript $\e\equiv\N$ to the
solution to indicate that it depends on the value of~$N$.
One can read off that the effective potential seen by eigenvalues $|x|{\gg}1$
approaches $V(x)-2\ln|x|$.  The situation is different, however, near
the minimum of the potential where $V'$ dominates the \lhs\ of eq.~\SPrhoc\
and most of~$\rh$ is concentrated.\foot{
The case of multiple local minima will be discussed in a while.}
When $\rh$ is ${\cal O}(1)$ the $\rh'\over\rh$-term may safely be neglected or
treated as a $\N$ perturbation in~\SPrhoc, leading to
$$
\rh_\e(x)\ \approx\ -{1\over\pi^2}-\!\!\!\!\!\!\int\!{dy\over x-y}\;\shalf
V'(y)
\eqn\mock2 $$
by simply inverting the Hilbert transform in equation~\SPrhoc.
The result is a modulation of Wigner's semicircle distribution.
When $N\to\infty$, the crossover regions between the interior and exterior
of the `Fermi sea' shrink to points $x{=}a,b$, and the saddle-point
equation~\SPrhoc\ turns into~[\brezin]
$$
\eqalign{
\shalf V'(x)\ &=\ -\!\!\!\!\!\!\int_a^b \!{dy\over x-y}\,\rh_0(y)
\ \equiv\ \pi\,\T_x[\rh_0] \qquad {\rm for}\ x\in[a,b]\cr
\rh_0(x)\ &=\ 0 \qquad\qquad\qquad\qquad\qquad\qquad\quad
{\rm for}\ x\notin[a,b] \quad.\cr}
\eqn\velocity $$
This relation is known as the {\it airfoil equation\/} and determines
the vorticity~$\rh_0$ related to a given velocity field~$V'$ along the
airfoil ($b\-a$ is the span of the wings)~[\sohngen].  In our case the
location of $a$ and~$b$ is determined from the normalization of~$\rh_0$.
Equation~\velocity\ is solved by inverting the {\it finite\/} Hilbert
transform~\T,
$$
\rh_0(x)\ =\ {1\over\sqrt{(b-x)(x-a)}}\biggl[{1\over\pi}\ -\ {1\over\pi^2}
-\!\!\!\!\!\!\int_a^b \!{dy\over x-y}\;\shalf V'(y)\;\sqrt{(b-y)(y-a)}\biggr]
\quad.\eqn\vorticity $$
Again, I have assumed a single-well potential~$V$, so that the support of
$\rh_0$ is a single, connected interval~$[a,b]$.

It is worthwhile to give the form of the saddle-point action.
Employing eq.~\SPlambda\ as well as $\,\ln\int e^{-\iN\lh}=-\iN\la_0\,$
I find, in agreement with ref.~[\menahem], that ($\rh\=\rh_\e$)
$$
\eqalign{
S_N[\rh,\lh]\ &=\
N^2\int\!\rh(x)V(x)-N^2\int\!\!\!\!\int\!\rh(x)\ln|x\-y|\rh(y)
+N\int\!\rh(x)\ln\rh(x)\cr
&=\ {\textstyle{N^2\over2}}\int\!\rh(x)(V(x)\-\ln|x|)
+{\textstyle{N^2\over2}}V(0)+{\textstyle{N\over2}}\int\!\rh(x)\ln\rh(x)
+{\textstyle{N\over2}}\ln\rh(0) \cr}
\eqn\SPaction $$
where I dropped a (singular) term $\shalf Nf(0)$ and made use of the
saddle-point equation~\SPrhoc.
The last integral permits an interpretation as the entropy of the
distribution~$\rh$.
Taking the naive large-$N$ limit, I obtain
$$
S_\infty[\rh_0]\ =\
{\textstyle{N^2\over2}}\int\!dx\;\rh_0(x)\,V(x)-N^2\ga\quad,
\eqn\Sinfty $$
with
$$
\ga\ =\ \int\!dy\;\ln|x-y|\,\rh_0(y) - \shalf V(x)\ =\
{\rm constant~~for~}x\in[a,b]
\eqn\chempot $$
being the chemical potential (or Lagrange multiplier enforcing $\int\rh_0~=1$).

At this point, I would like to drive home an essential point of my talk.
The non-trivial $N$-dependence of the classical backgound~$\rh_\e$ implies
that the classical limit, $\r\to\rh_\e$, {\it differs\/} from the
low-temperature limit, $N\to\infty$, because of $\N$ corrections coming from
the integration measure.
As a consequence, the semiclassical loop expansion will {\it not\/} be
identical
to the topological $\N$ expansion; rather, a {\it double\/} expansion arises.
Beyond this, non-perturbative (in $\N$) contributions appear already at tree
level (in $\hbar$).
It is therefore by no means clear that interchanging limits, by {\it first\/}
taking $N\to\infty$ in equation~\SPrhoc\ and {\it then\/} solving
equation~\velocity\ to obtain $\rh_0$, provides the large-$N$ limit,
$\lim_{\e\to0}\rh_\e$, of a proper solution to equation~\SPrhoc,
as my notation suggests.
In fact, I will now demonstrate that in general a solution $\rh_0$ does
{\it not\/} correspond to a finite-$N$ saddle point~$\rh_\e$.

Non-perturbative effects become tangible when two or more potential wells
compete for eigenvalues.
For even potentials, the full complexity of the problem appears first in the
triple-well potential
$$
V(x)\ =\ \shalf x^2 + g x^4 + \e x^6 \quad,\qquad\e>0\quad,
\eqn\pot $$
since the $x\leftrightarrow-x$ symmetry trivializes the double-well case.
I like to fix $\e$ to some small value and probe the phase diagram by
decreasing
the quartic coupling~$g$ along the negative axis.
For $g<g_*=-\sqrt{3\e/2}$ the potential develops three well-separated minima
which become degenerate at $g=g_@=-\sqrt{2\e}$.
For sufficiently negative~$g$ the `Fermi sea' must, therefore,
consist of three or two disconnected oceans, called {\it arcs\/} or {\it
bands}.
At $N\to\infty$, the eigenvalue density should then be supported on one, two,
or three disjoint intervals.
As shown in refs.~[\bhanot,\jain,\ray], the general solution of
equation~\velocity\ reads
$$
\rh_0(x)\ =\ {\textstyle{3\e\over\pi}}\,(n^2-x^2)\,\sqrt{
(4a^2-x^2)\,(4b^2-x^2)\,(4c^2-x^2)} \quad,
\eqn\rhothree $$
describing a positive three-band density with
$0\le2a<n<2b\le2c$ and support on $[-2c,-2b]\cup[-2a,2a]\cup[2b,2c]$.
Normalization imposes three conditions on $\{n,a,b,c\}$ which leaves a
one-parameter {\it family\/} of solutions.
A convenient parameter to label these solutions is the difference
$\D\ga=\ga_o-\ga_c$
of the chemical potentials~$\ga$ for the outer and the central bands.
The chemical potential is nothing but the integration constant appearing when
integrating equation~\velocity. Since it may take different constant values for
two eigenvalues $x_c\in[-2a,2a]$ and $x_o\in[2b,2c]$ from two different bands,
the difference
$$
\D\ga\ =\ \int\! dy\;\ln{|x_o\-y|\over|x_c\-y|}\,\rh_0(y)
- \shalf\bigl[V(x_o)\-V(x_c)\bigr]
\eqn\chempotdiff $$
is a genuine property of the solution~$\rh_0$.
In this situation one must replace
$$
\ga\longrightarrow\sum_{\rm bands}\ga_i n_i\quad,\qquad
n_i=\int_{i\rm th~band}\!\!\!\!\!\!dx\;\rh_0(x)\quad,\qquad
\sum_{\rm bands}n_i=1\quad,
\eqn\replace $$
in equation~\Sinfty.
Extremal values of~$\D\ga$ occur when the number of bands decreases, for
$b\to c$ or $a\to0$, and the density~\rhothree\ becomes unique.
However, those solutions do not exist everywhere in the $(\e,g)$ plane.
For sufficiently small $\e$, the range of one-, two- and three-band solutions
is given by the sequence $g_3<g_@<g_2<g_*<g_1<0$.
Here, {\it one-\/}band densities arise for $g>g_3$,~\foot{
$g\=g_3(\e)$ is the critical BPIZ line where the double scaling limit
is to be taken~[\brezin].}
{\it two-\/}band distributions occur for $g<g_2$, and
{\it three-\/}band solutions appear for $g<g_1$.
Hence, equation~\velocity\ admits a {\it unique\/} solution only for $g>g_1$.
The overlapping regions above indicate
a coexistence of multiple-band densities elsewhere.
Even more astonishing is the discovery that the members~$\rh_0(x,\D\ga)$
of such a family are {\it not\/} degenerate in free energy~[\ray].
This contradiction in terms is resolved by noticing that infinitesimal
variations within the family, which correspond to the tunneling of individual
eigenvalues, are actually {\it singular\/} at the band edges.
One might say that it requires a {\it finite\/} variation to move an
individual eigenvalue to another well although it is only an ${\cal O}(\N)$
effect.
Since ${\partial\over\partial\D\ga}\rh_0$ is not square-integrable
the inclusion of this mode among the density fluctuations in debatable.
As the chemical potential drives the tunneling of eigenvalues,
one should expect the {\it minimal\/} action to belong to the unique family
member with $\D\ga\=0$, which I call {\it dominant\/}.
This is indeed what happens and can be checked numerically~[\ray].

Of course, for $N<\infty$ there is always a {\it unique\/} saddle-point density
$\rh_\e$, with a unique limit as $\e\to0$, because the strict positivity of the
distribution implies that $\D\ga\=0$ all along.
Hence, we have
$$
\lim_{N\to\infty} \rh_\e(x)\ =\ \rh_0(x,\D\ga\=0)\quad.
\eqn\stable $$
With hindsight it is clear that the `sub-dominant' members of a three-band
family could only appear because I took the limit $N\to\infty$ prematurely
by going from equation~\SPrhoc\ to~\velocity.
`Physically' speaking, the freezing of the Dyson gas of eigenvalues at
{\it zero\/} temperature entirely suppresses any tunneling and permits those
ficticious saddle-point distributions.
At {\it finite\/} temperature, the Dyson solid melts at the edges,
and tunneling, although exponentially small, destabilizes all but the
dominant solutions.
The `entropy term' in equation~\SPrhoc\ plays the crucial role.
However, I may still employ the incomplete large-$N$ saddle-point
equation~\velocity, if it is complemented by the additional
global requirement $\D\ga\=0$.

In this light the correct phase diagram for the potential~\pot\ looks as
follows.  Again taking $\e$ small but fixed, I find a new sequence
$g''<g_@<g'<0$, where a small region of three-band dominance around the
degenerate point ($g\=g_@$) separates two-band from one-band densities.
Inspecting the shape of the potential at the transition values $g'$ and $g''$,
one learns that essentially the number of {\it degenerate absolute minima\/}
determines the number of eigenvalue bands.
Interestingly, the jump from one to three well-separated bands is smoothed out
by an interpolating three-band region occuring when non-degenerate wells are in
some sense comparable and can compete for eigenvalues.
\vglue.2in
\goodbreak

\noindent
{\bf{Orthogonal Polynomials.}}

\noindent
The standard approach to matrix model calculations, and so far the only one
capable of producing the topological expansion and the double-scaling limit,
is the method of orthogonal polynomials~[\bessis].
Its starting observation is that the van der Monde determinant in
$$
e^{-N^2 F_N}\ =\ \biggl[\prod_{i=1}^N \int\! dx_i\biggr] \;
\prod_{i<j} (x_i-x_j)^2 e^{-N\sum_i V(x_i)}
\eqn\Fxdef $$
can be rewritten as
$$
\prod_{i<j} |x_i-x_j|\ =\ \left|\matrix{
1&1&\ldots&1\cr x_1&x_2&\ldots&x_N\cr x_1^2&x_2^2&\ldots&x_N^2\cr
\vdots&\vdots&\ddots&\vdots\cr x_1^{N{-}1}&x_2^{N{-}1}&\ldots&x_N^{N{-}1}\cr}
\right| \ =\ \left|\matrix{
P_0(x_1)&P_0(x_2)&\ldots&P_0(x_N)\cr P_1(x_1)&P_1(x_2)&\ldots&P_1(x_N)\cr
P_2(x_1)&P_2(x_2)&\ldots&P_2(x_N)\cr \vdots&\vdots&\ddots&\vdots\cr
P_{N{-}1}(x_1)&P_{N{-}1}(x_2)&\ldots&P_{N{-}1}(x_N)\cr} \right| \quad,
\eqn\vandermonde $$
with monic polynomials $P_k(x)=x^k+{\rm (lower~order)}$.
If one cleverly chooses the $P_k$ to be mutually orthogonal with respect to the
measure $e^{-NV(x)}$,
$$
h_k\,\d_{kl}\ =\ \int\!dx\;e^{-NV(x)}\;P_k(x)\,P_l(x)\quad,
\eqn\Pdef $$
the change of basis $x^k\to P_k(x)$ in equation~\Fxdef\ exactly diagonalizes
the Coulomb interaction and trivializes the free energy to
$$
F_N\ =\ -{\textstyle{1\over N^2}}\,\ln[N!h_0h_1h_2\ldots h_{N{-}1}]\quad.
\eqn\FN $$
The construction of the polynomials simplifies thanks to the classic
recursion relation
$$
P_{k+1}(x)\ =\ x\,P_k(x) - R_k\;P_{k-1}(x) \quad,\qquad
R_k\ =\ {h_k\over h_{k-1}}\ \ge 0\quad,
\eqn\Prec $$
so that is it sufficient to compute the norms $h_k$, starting with the initial
condition $R_0\=0$.

The Stieltjes method of constructing the orthogonal polynomials consists of
iterating eq.~\Prec\ and the norm computation,~\Pdef:
$$
P_k(x;R_0,\ldots,R_{k{-}1})\longrightarrow h_k \longrightarrow R_k
\longrightarrow P_{k{+}1}(x;R_0,\ldots,R_k) \quad.
\eqn\stieltjes $$
Unfortunately, it is quite inappropriate for numerical analysis.
However, the special form $e^{-NV}$ of the measure allows for a finite
recursion relation among the $R_k$ themselves,
the `string equation'~[\bessis]
$$
\eqalign{
{\textstyle{k\over N}}\ =\ R_k\,\Bigl\{1&+4g\bigl(R_{k-1}+R_k+R_{k+1}\bigr)+
6\e\bigl(R_{k-1}+R_k+R_{k+1}\bigr)^2 \cr
&+6\e\bigl(R_{k-2}R_{k-1}-R_{k-1}R_{k+1}+R_{k+1}R_{k+2}\bigr)\Bigr\}\quad,\cr}
\eqn\Rrec $$
displayed here for the potential~\pot.
After solving for $\ R_k=R_k(R_{k-1},\ldots,R_{k-4})\ $ one still needs
the initial values
$$
R_1\ =\ {h_1\over h_0}\ =\ {\int e^{-NV}\,x^2\over\int e^{-NV}}\quad,\qquad
R_2\ =\ {h_2\over h_1}\ =\
{\int e^{-NV}\,(x^2\-R_1)^2\over\int e^{-NV}\,x^2}
\eqn\Ronetwo $$
besides $\ R_{-1}=0=R_0\ $ to start the iteration.
It turns out that this procedure is numerically tractable, but
$R_k$ becomes increasingly sensitive to the initial conditions for growing
$k$ or~$N$~[\tunneling].
Trivial but instructive is the exactly solvable example of purely quadratic
potential, i.e. $g\=\e\=0$, or $V\=\shalf x^2$.
In this case one simply rediscovers the Hermite polynomials from $R_k=k/N$,
as appropriate for the harmonic oscillator.

Let us now investigate the large-$N$ or planar limit, in order to connect up
with the semiclassical results.
Here, I have to rely on an assumption, namely
a continuum approach of the $R_k$ needs to be {\it postulated}.
Writing
$$
{\textstyle{k\over N}}\ =\ \x\in[0,1]\quad,
\qquad R_k\ =\ r_\e(\x)\quad,\quad \e=\N\quad,
\eqn\condnot $$
the simplest ansatz
$$
R_{k{+}1}-R_k\ =\ {\cal O}(\N) \qquad {\rm as} \quad N\to\infty
\eqn\oneRansatz $$
implies $r_\e(\x)\to r(\x)$, a smooth positive function with $r(0)\=0$.
Numerical studies show that this behavior indeed occurs whenever a single
potential well is clearly dominant~[\tunneling].
This coincides with the one-band regime of our potential.\foot{
excluding regions of ill-separated minima.}
The string equation is then dramatically simplified to the algebraic
relation~[\bessis]
$$
\x\ =\ r\bigl\{ 1 + 12gr + 60\e r^2 \bigr\} \quad.
\eqn\oner $$
The condition $r(0)\=0$ selects a unique branch of~$r$ which monotonically
reaches $\x\=1$ provided $g>g_3(\e)$, the BPIZ critical line.
Finally, the free energy $F_\infty$ is obtained by
naively taking the $N\to\infty$ limit of eq.~\FN,
$$
\eqalign{
F_N\ &=\ -{\textstyle{1\over N^2}}\ln N! - \N\ln h_0
- \N\sum_{k=1}^N\left(1-{\textstyle{k\over N}}\right)\ln R_k \cr
&\rightarrow\ -{\textstyle{1\over N^2}}\ln N! + V_{\rm min}
- \int_0^1\!d\x\;(1-\x)\,\ln\r(\x) \quad,\cr}
\eqn\Finfty $$
to be compared to ${1\over N^2}S_\infty$ from equation~\Sinfty.

However, the continuity assumption~\oneRansatz\ is clearly violated when
$g<g_@$, because an estimate of $R_1$ from equation~\Ronetwo\ reveals that
it must {\it jump\/} from ${\cal O}(\N)$ to the square of the location of the
outer potential minima, $\approx{-}{g_@\over3\e}\={1\over\sqrt{6\e}}$,
when $g$ drops below the degenerate point, $g\=g_@$.
Such a behavior is known from studies of double-well
potentials~[\molinari,\jain], where an {\it alternating\/} sequence
$$
R_{k{+}2}-R_k\ =\ {\cal O}(\N)\quad,\qquad
r_\e(\x)\ \to\ \cases{r^{(0)}(\x)&for $k$ even\cr r^{(1)}(\x)&for $k$ odd\cr}
\eqn\twoRansatz $$
is observed.
Under this modified assumption, the string equation~\Rrec\ turns into two
coupled cubic equations for $r^{(0)}$ and $r^{(1)}$.
Their graphical solution exhibits $r^{(0)}(\x)$ as an increasing function
starting from $r^{(0)}(0)\=0$, and $r^{(1)}(\x)$ as decreasing from
$r^{(1)}(0)=x_{\rm min}^2$~[\ray].

Like for the saddle-point method, it is not evident which large-$N$ assumption
is to be chosen for given values of $\e$ and~$g$.
It can be shown, however, that the series of $\sqrt{R_k}$ approximates the
sequence of consecutive eigenvalues $x_k$ eventually building up to the
distribution $\rh_\e(x)$ in the limit $\e\to0$.
Hence, one- and two-band regions in the phase diagram must correspond to
the continuum behavior of eqs. \oneRansatz\ and~\twoRansatz,
respectively.~\foot{
Actually, the number of branches $r^{(i)}$ reaching the point $\x\=1$ is
relevant.}
An unsettling gap remains, however, in the three-band dominated buffer zone.
How can the one-branch ansatz~\oneRansatz\ merge with the two-branch
ansatz~\twoRansatz\ when
$r^{(1)}(0)-r^{(0)}(0)\approx x_{\rm min}^2\geq{1\over\sqrt{6\e}}$?
The resolution of this paradox was discovered through a numerical
analysis, which uncovered a $\x$~interval $[\bar{\x},\tilde{\x}]$ with
seemingly
{\it chaotic\/} recursion coefficients~$R_k$, interpolating between
one or two branches for $\x<\bar{\x}$
and a single branch for $\x>\tilde{\x}(>1)$.
The three-band region $g''<g<g'$ coincides with $\bar{\x}<1$, i.e. the
onset of the unpredictable behavior creeping into the relevant
$\x$~interval~$[0,1]$.
Only exactly at the degenerate point, $g\=g_@$, is a simple three-branch
solution realized~[\ray]. This picture has been confirmed by several
groups~[\suzuki,\senechal].
\vglue.2in
\goodbreak

\noindent
{\bf{Conclusions.}}

\noindent
I have reviewed some non-perturbative aspects of hermitean random matrix
models, with an emphasis on the distribution of eigenvalues among several
potential wells.
It turned out that the classical, quasi-continuous, large-$N$ eigenvalue
distribution depends {\it non-perturbatively\/} on~$\N$, so that the
semiclassical loop expansion must be distinguished from the standard
topological (or string loop) expansion.

For multiple-well matrix potentials, interchanging the limits $N{\to}\infty$
and
$S_N{\to}$extremum is dangerous due to the $N{<}\infty$ equilibration between
different Dyson gas components, mediated by eigenvalue tunneling.
As a consequence, the point of critical coupling becomes {\it metastable\/}
for pure gravity when $V$ gets bounded from below.

{}From numerical simulations of the sequence of recursion coefficients~$R_k$
for
the orthogonal polynomials, I conjecture that their behavior is characterized
by the critical points of the potential function~$V$ itself.
The consecutive equilibrium deposition of eigenvalues on the real line is
suggested to be sensitive to features of~$V$ at increasing values.
For {\it degenerate\/} absolute minima one observes a quasi-periodic series
of~$R_k$, whereas {\it non-degenerate\/} minima produce {\it chaotic\/}
behavior! The unpredictability of the coefficients reflects the competition
of incommensurate potential wells for eigenvalues.

I hope to have demonstrated that the continuum limit of matrix models is more
complicated than imagined originally.
In view of this it would be very beneficial to understand their critical
properties in the semiclassical description.
An attempt to push the latter beyond the classical limit is currently in
progress (see~[\semicl]).
\ack
I am grateful to the organizers for the charming and stimulating atmosphere
of the Symposium, and especially K. Behrndt for his personal efforts to make
it all work.
%
\refout
\bye